\documentclass[aip,apl,amsmath,amssymb,reprint,]{revtex4-2}

\usepackage{graphicx}
\usepackage{dcolumn}
\usepackage{bm}
\usepackage{tabularx}
\newcolumntype{C}{>{\centering\arraybackslash}X}
\usepackage{multirow}
\usepackage[font={footnotesize}]{caption}
\usepackage{soul}
\usepackage[normalem]{ulem}
\usepackage[dvipsnames]{xcolor}
\usepackage{pdfsync}
\usepackage{soul}
\usepackage{braket}

\usepackage{color}
\newcommand{\mjm}[1]{\textcolor{black}{#1}}
\newcommand{\dds}[1]{\textcolor{black}{#1}}
\newcommand{\peb}[1]{\textcolor{black}{#1}}
\newcommand{\markup}[1]{\textcolor{black}{#1}}

\begin{document}

\author{David P. Lake}

\author{Matthew Mitchell}

\author{Denis D. Sukachev}

\author{Paul. E. Barclay}
\affiliation{Department of Physics and Astronomy and Institute for Quantum Science and Technology, University of Calgary, Calgary, AB, T2N 1N4, Canada}
\email{pbarclay@ucalgary.ca}

\date{\today}

\title{Processing light with an optically tunable mechanical memory}

\begin{abstract}
\peb{Mechanical systems are one of the promising platforms for classical and quantum information processing and are already widely-used in electronics and photonics. Cavity optomechanics offers many new possibilities for information processing using mechanical degrees of freedom; one of them is storing optical signals in long-lived mechanical vibrations by means of optomechanically induced transparency.} However, the memory storage time is limited by intrinsic mechanical dissipation. More over, in-situ control  and manipulation of the stored signals--processing--has not been demonstrated. Here, we address both of these limitations using a multi-mode cavity optomechanical memory. An additional optical field coupled to the memory modifies its dynamics through time-varying parametric feedback. We demonstrate that this can extend the memory \peb{decay} time by an order of magnitude, decrease its effective mechanical dissipation rate by two orders of magnitude, and deterministically shift the phase of a stored field by over 2$\pi$. \peb{This further expands the information processing toolkit provided by cavity optomechanics.}
\end{abstract}

\maketitle

Information processing devices exhibit dissipation due to their coupling to external degrees of freedom. This permits energy to leave and fluctuations to enter the system \cite{ref:callen1951iag}, and  typically degrades {the performance of components such as memories}. \peb{ Within the field of optomechanics \cite{ref:aspelmeyer2014co, ref:safavi-naeini2019cpp}, this has motivated tremendous progress in reducing intrinsic dissipation through precise tailoring of device geometry and materials \cite{ref:ghadimi2018ese, ref:maccabe2019pbn, ref:otterstrom2018sbl, ref:tsaturyan2017unr}. A complementary approach, sometimes referred to as { reservoir engineering\cite{ref:poyatos1996QRE}},}  uses dissipation \peb{channels} to enhance system properties, \peb{ typically via  \peb{an auxiliary} coherent source that couples to the system.} Here we show that when reservoir engineering is \dds{generalized}  to incorporate dynamic control of this external coupling, a system's 	 steady state can be adiabatically manipulated. \peb{ By dynamically varying the coupling of an optomechanical memory to an auxiliary control field--a reservoir mode--}we demonstrate that stored light can be coherently modified. This is a crucial step towards realizing \peb{\dds{optical processing}  of light stored in the mechanical motion of a nanoscale device.}

\begin{figure}[ht]
\centering
\includegraphics[width=0.95\linewidth]{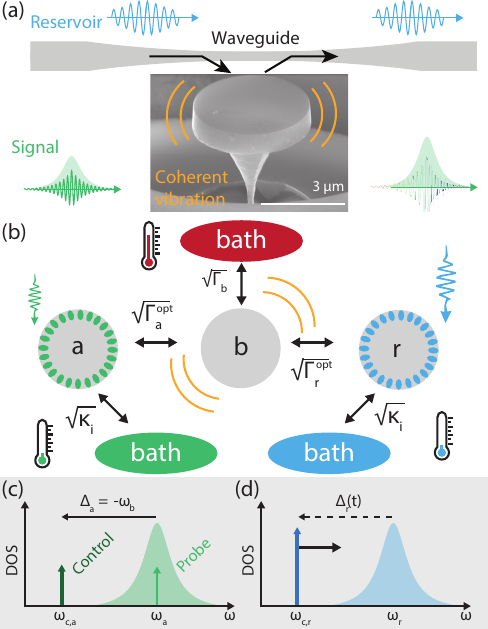}
\caption{{\bf Key elements of the tunable optomechanical memory} (a) Annotated scanning electron micrograph of the diamond microdisk optomechanical cavity. A tapered optical fiber is utilized to couple light into two of the microdisk \dds{whispering gallery modes} which in turn  \dds{are} coupled to mechanical vibrations of the radial breathing mode of the microdisk. (b) Schematic of the system under study where the mechanical mode $b$ is coupled to the environment (red bath), an \peb{optical} reservoir mode, $r$, and the optical mode $a$ used for writing and reading information, at the indicated rates. While coupling of the mechanical mode to the surrounding environment is an intrinsic property of the device and environment, the coupling to $r$ may be manipulated through the use of a control laser. (c,d) Density of states (DOS) picture showing the detuning of the control and probe laser fields for (c) mode $a$ and (d) mode $r$. \peb{ Here $\omega_\text{c,a}$ and $\omega_\text{c,r}$ are the frequencies of the control lasers for the signal and reservoir modes, respectively.}}
\label{fig:schematic}
\end{figure}

\peb{ The concept of using an external field to manipulate optical information via coupling to a mechanical degree of freedom has been previously explored in Brillouin scattering optomechanics \cite{ref:eggleton2019bip}, where powerful functionality including pulse storage \cite{ref:zhu2007sli, ref:merklein2017aci}, all optical signal processing \cite{ref:santagiustina2013asp}, and coherently refreshed memory \cite{ref:stiller2019cra} have been demonstrated in  cm-long waveguides.  Cavity optomechanical devices, in which optical and mechanical modes can be confined and spatially overlapped within wavelength-scale volumes, allow nearly complete transfer of an optical excitation to low dissipation  mechanical resonances. These devices operate with relatively low power and occupy micron-scale footprints, and have been used for information storage \cite{ref:fiore2011soo, ref:walluks2019Memory}. However, manipulating this stored information has not been demonstrated.}

\peb{In this work,  we use a diamond optomechanical microdisk cavity to realize a memory based on optomechanically-induced transparency \cite{ref:weis2010oit, ref:safavi2011eit} that converts a weak optical input to a long-lived mechanical excitation. We then show that by adjusting the frequency of a field input to a reservoir mode we can effectively reduce the mechanical dissipation of the device through parametric feedback until it is just below self-oscillation threshold, allowing the memory lifetime to be extended.  Through precise tuning of the reservoir field frequency, we directly observe a memory lifetime enhancement of over 7 times, and realize an over 150 times reduction in the resonator's effective mechanical dissipation rate. Finally, we demonstrate that by dynamically varying the reservoir field frequency the phase of the stored signal can be manipulated.}

\section{Results}

Diamond microdisks can operate as multimode cavity optomechanical systems whose optical whispering gallery modes are coupled by radiation pressure to motion of the device's mechanical resonances \cite{ref:mitchell2016scd}. \peb{ Coherent multimode optomechanical coupling is possible in these devices even at room temperature and ambient conditions \cite{ref:lake2018oit} thanks to the low energy dissipation rates, $\kappa$ and $\Gamma$, of their optical and mechanical modes, respectively, combined with diamond's ability to support high intensity fields without suffering from nonlinear absorption and heating. This latter property increases the maximum number ($n$) of photons that can be supported by the microdisk, which enhances the photon-assisted optomechanical coupling rate $g = \sqrt{n} g_0$. These devices have a high single-photon optomechanical coupling rate $g_0$ due to their wavelength-scale dimensions and strong spatial overlap of the microdisk's whispering gallery modes with its mechanical radial breathing mode.}

\peb{The condition for coherent optomechanical coupling in both classical and quantum devices is cooperativity  $C = 4g^2/\kappa\Gamma > 1$, and \markup{can be achieved} simultaneously by multiple modes of a diamond microdisk.} Multimode cavity optomechanics enables \dds{optical} wavelength  conversion \cite{ref:hill2012cow, ref:dong2012ODM, ref:liu2013eit, ref:mitchell2019oaw}, \dds{photon entanglement} \cite{ref:barzanjeh2018ser, ref:chen2019eop}, and low-noise frequency conversion \dds{in the microwave domain} \cite{ref:ockeloen2016LNA}. We  show \dds{in this paper} that multimode diamond microdisks are an excellent platforms for implementing \peb{memories whose stored information can be manipulated via dynamic reservoir engineering} \cite{ref:toth2017adq, ref:fang2017gnr}.

The system used in this work is illustrated schematically in Fig.~\ref{fig:schematic}.  Information input to an optical `signal' mode ($a$) of a diamond microdisk cavity is transferred via optomechanical coupling to the device's mechanical mode ($b$). Simultaneously,  the mechanical mode  dynamics are modified through its coupling to an optical  `reservoir'  mode ($r$). The optical and mechanical modes are characterized by their frequencies ($\omega_\mathrm{a,r}$, $\omega_\mathrm{b}$) and energy decay rates ($\kappa_\mathrm{a,r}$, $\Gamma_\mathrm{b}$), respectively. The reservoir mode is driven by a control laser whose detuning from resonance, $\Delta_\mathrm{r}$, sets the phase lag of its  optomechanical coupling to the mechanical resonator, and whose power, $P_\mathrm{r}$ (defined here as the power input to the fiber taper waveguide), sets the coupling strength. This tunable resonator--reservoir interaction induces additional mechanical dissipation $\Gamma_\mathrm{r}^\mathrm{opt}$ and shifts the mechanical resonator frequency by $\omega_\mathrm{r}^\mathrm{opt}$, two effects widely studied in single--mode optomechanical systems, for example in demonstration of mechanical ground state cooling \cite{ref:chan2011lcn, ref:teufel2011scm}.

\peb{To process information stored in the mechanical resonator, we dynamically adjust the power and  detuning of the
reservoir input field}. The resonator evolution \dds{is governed by (Supplementary Material):}
\begin{align}\label{eq:mechReservoir}
\dot{\hat{b}} =&\ -\left(i\omega_\mathrm{b}^\mathrm{eff}(\Delta_r, g_r) + \frac{\Gamma_\mathrm{b}^\mathrm{eff}(\Delta_r, g_r)}{2} \right)\hat{b} +\sqrt{\Gamma_\mathrm{b}}\hat{e}_\mathrm{in} \nonumber\\
&\ +g_\mathrm{r} \sqrt{\kappa_\mathrm{r}}\chi_\mathrm{r}(\omega_\mathrm{b};\Delta_r)\hat{r}_\mathrm{in} + g_\mathrm{r} \sqrt{\kappa_\mathrm{r}}\chi_{\mathrm{r}^\dagger}(\omega_\mathrm{b};\Delta_r)\hat{r}_\mathrm{in}^\dagger,
\end{align}
where $\hat{b}$ is the phonon annihilation operator, $\hat{e}_\mathrm{in}$ is the thermal bath input field, $\hat{r}_\mathrm{in}$ is the optical reservoir input field,  $g_\mathrm{r} \propto \sqrt{P_r}$ is the photon-assisted optomechanical coupling rate between the reservoir mode and the resonator, and $\chi_\mathrm{r}$ is the reservoir mode's optical response in the frame of the reservoir mode control laser. The key feature that we test and exploit is the ability to dynamically control the memory's effective mechanical frequency, $\omega_\mathrm{b}^\mathrm{eff}=\omega_\mathrm{b}+\omega_\mathrm{r}^\mathrm{opt}(t)$, and effective damping rate, $\Gamma_\mathrm{b}^\mathrm{eff}=\Gamma_\mathrm{b}+\Gamma_\mathrm{r}^\mathrm{opt}(t)$, \peb{by adjusting the reservoir mode parameters $g_r(t)$ and $\Delta_r(t)$ as a function of time}. Notably, we find that when we input a field to mode $a$ to this system, the memory operates {{as if}} it is a conventional \peb{single-mode} cavity optomechanical device whose mechanical resonator dynamics have been renormalized. This regime is valid if $\Gamma_\mathrm{b} \ll \kappa_\mathrm{r}$.

\peb{
Equation  \eqref{eq:mechReservoir}  shows that the reservoir mode coupling acts as a dissipative process, and not a direct drive. Since we alter the effective dissipation of the system, we use the descriptor { reservoir engineering}, in analogy to the work that introduced this term\cite{ref:poyatos1996QRE}.}
\peb{Note that although the effects demonstrated below \dds{can be described classically}, the quantum \dds{formalism} used here allows the analysis of noise processes, including photon-phonon scattering, that will ultimately limit memory performance.}

Below we test the validity of this description and demonstrate applications of dynamic reservoir coupling through three experiments. First, we \dds{show an enhancement of} the system's optomechanical cooperativity due to its renormalized mechanical dissipation, and switch the dynamics of the system from overall loss to overall gain. Second, we \dds{increase} the optomechanical memory storage time through control of $\Gamma_\mathrm{b}^\mathrm{eff}$. Finally, we \dds{control the phase of a stored mechanical signal through manipulation of $\omega_\mathrm{b}^\mathrm{opt}$ as a function of time.
}

\subsection{Engineering the system dynamics}

\begin{figure}
\centering
\includegraphics[width=\linewidth]{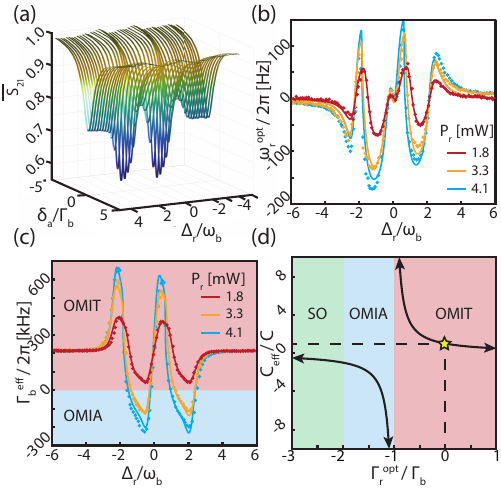}
\caption{{\bf Tuning the mechanical resonator dynamics} (a) Normalized OMIT scans as a function of mode $a$ probe-cavity detuning, $\delta_\mathrm{a}$, and control-reservoir detuning, $\Delta_\mathrm{r}$, and $S_{21}$ is the probe laser reflection measured using a vector network analyzer. The changes in the transparency window as a function of $\Delta_\mathrm{r}$ are indicative of reservoir interactions. Optomechanically manipulated effective  (b) mechanical frequency and (c) damping as a function of $\Delta_\mathrm{r}$ and the reservoir mode input power, $P_\mathrm{r}$. (d) Illustration of effective cooperativity, $C_\mathrm{eff}$, for varying $\Gamma_{\mathrm{b}}^\mathrm{eff}$ controlled by the reservoir mode, when $\Gamma_{\mathrm{a}}^\mathrm{opt} = \Gamma_\mathrm{b}$ due to the presence of the mode $a$ control field. Three different regimes of operation are shown: OMIT, OMIA, and self-oscillation (SO). The yellow star indicates the operating point when the reservoir field is off.}
\label{fig:1}
\end{figure}

We first probe how the dynamics of the mechanical resonator, and its resulting coupling to light, are affected by the resonator-reservoir interaction. This is accomplished using optomechanically induced transparency (OMIT) spectroscopy \cite{ref:weis2010oit, ref:safavi2011eit}.   OMIT creates a transparency window in the cavity lineshape whose properties depend on the dynamics of the optomechanical system. By coherently coupling a probe field in $a$ to the mechanical resonator for varying reservoir control laser settings,  we can learn about the influence of the reservoir on the resonator. These measurements are shown in Fig.~\ref{fig:1}(a), which were obtained using a fiber taper waveguide to evanescently couple the reservoir control laser to mode $r$ ($\omega_\mathrm{r}/2\pi = 192~\mathrm{THz},~\kappa_\mathrm{r}/2\pi=1.13~\mathrm{GHz}$) for varying $\Delta_\mathrm{r}$, while performing OMIT spectroscopy on mode $a$ ($\omega_\mathrm{a}/2\pi = 197~\mathrm{THz},~\kappa_\mathrm{a}/2\pi=0.856~\mathrm{GHz}$) using a weak resonant probe laser and a control laser red detuned by $\omega_\mathrm{b}$. Note that in all measurements presented below the probe field is generated by modulating the control laser and is typically in the $\mu$W range (see Methods). The device used here has $\omega_\text{b}/2\pi \sim 2.14\,\text{GHz}$ and $\Gamma_\text{b}/2\pi =  190$ kHz \peb{ for the radial breathing mode} and operates in the resolved sideband regime for modes $a$ and $r$ ($\omega_\mathrm{b}/\kappa_\mathrm{a} = 2.5,~ \omega_\mathrm{b}/\kappa_\mathrm{r} = 1.9)$. \peb{Coupling to other mechanical modes of the microdisk was not observed.}  Note that the optical reservoir mode is a standing wave doublet (see Methods and Supplementary Material) formed from backscattering in the microdisk \cite{ref:kippenberg2002mct,ref:borselli2005brs}, whose most apparent effect is the two sets of minima and maxima in Fig.~\ref{fig:1}(a). In all measurements the mode $a$ control field detuning was set relative to the lowest frequency doublet resonance.

This measurement was repeated for three different values of $P_\mathrm{r}$, with the probe control laser fixed at  input power $P_\text{a} \sim 0.26$ mW (intracavity photon number $n_\text{a} \sim 3.2 \times 10^4$).  At each reservoir setting $\Gamma_\mathrm{b}^\mathrm{eff}$ and $\omega_\mathrm{b}^\mathrm{eff}$ were extracted from the OMIT window shape (see Supplementary Material), and are plotted along with fits to the data in Figs.\,\ref{fig:1}(b) and (c). The fits, which show excellent agreement with measurements,  were obtained with  $g_\mathrm{r}$ as the only fitting parameter.
\dds{This confirms that optomechanical renormalization of the mechanical resonator dynamics by the reservoir field affect mode $a$'s  optomechanical response as if  $\Gamma_\mathrm{b}^\mathrm{eff}$ and $\omega_\mathrm{b}^\mathrm{eff}$  were the mechanical resonator's intrinsic mechanical properties.}

The system dynamics are most dramatically affected when $\Gamma_\mathrm{b}^\mathrm{eff}$ approaches zero and \peb{ becomes negative}. In a conventional OMIT system, the depth and width of the transparency window is parameterized by the optomechanical cooperativity, $C = \frac{4|g_\mathrm{a}|^2}{\kappa_\mathrm{a}\Gamma_\mathrm{b}} = n_\text{a} \frac{4|g_0|^2}{\kappa_\mathrm{a}\Gamma_\mathrm{b}}$, \dds{where $g_\mathrm{a}$ is the photon-assisted optomechanical coupling rate for \dds{mode} $a$. For the microdisk used here, $g_0/2\pi \sim 25$\,kHz \cite{ref:mitchell2019oaw}. Note that the optomechanical cooperativity differs from the single-photon cooperativity by a factor of $n_\text{a}$.}
 However, in our multimode system OMIT is governed by an effective cooperativity $C_\mathrm{eff} = C \times \Gamma_\mathrm{b}/\Gamma_\mathrm{b}^\mathrm{eff}$. In the measurement presented here we achieve a maximum $C_\mathrm{eff} = 83$, which represents an enhancement of $158 \times$ the bare $C = 0.52$ \peb{experienced by the mode $a$ probe in absence of the reservoir field}.  This allows our system to act as though it has large cooperativity, enabling large light delays and narrow transparency windows (see Supplementary Material). {This enhancement is limited only by how close to zero $\Gamma_\mathrm{b}^\mathrm{eff}$ can be tuned through adjustment of $\Delta_\mathrm{r}$. Hence, it is not limited by available laser power since the regime of $\Gamma_\mathrm{b}^\mathrm{eff} < 0$ can be reached for the powers used here (see below)}.

As $\Gamma_\mathrm{b}^\mathrm{eff}$ becomes negative, the mechanical resonator motion changes from experiencing an overall loss to an overall gain. Consequently, by adjusting our reservoir coupling, we are able to tune the system dynamics between OMIT and the regime of optomechanically induced amplification (OMIA). This gain would normally cause optomechanically induced self-oscillation to occur \cite{ref:rokhsari2005rpd, ref:poot2012bls}. However, this instability is repressed in our multimode measurements by the optomechanical damping $\Gamma_\mathrm{a}^\mathrm{opt}$ induced by  mode $a$'s OMIT process, provided $\Gamma_\mathrm{a}^\mathrm{opt} + \Gamma_\mathrm{b}^\mathrm{eff} > 0$. The ability of the reservoir mode to tune the system dynamics between OMIT, OMIA and self-oscillation regimes is illustrated in Fig.~\ref{fig:1}(d) for the special case that $\Gamma_\mathrm{a}^\mathrm{opt} = \Gamma_\mathrm{b}$. In our measurements, the OMIA regime was entered for both the $P_\mathrm{r} = 3.3$ mW and $P_\mathrm{r} = 4.1$ mW settings when the control laser was blue detuned from the reservoir mode's  lowest frequency doublet feature by approximately $\omega_\mathrm{b}$. This regime was not accessed during the  storage measurements described in the following sections, and can only be measured due to the damping provided by the mode $a$ fields.

\begin{figure*}
\centering
\includegraphics[width=0.85\linewidth]{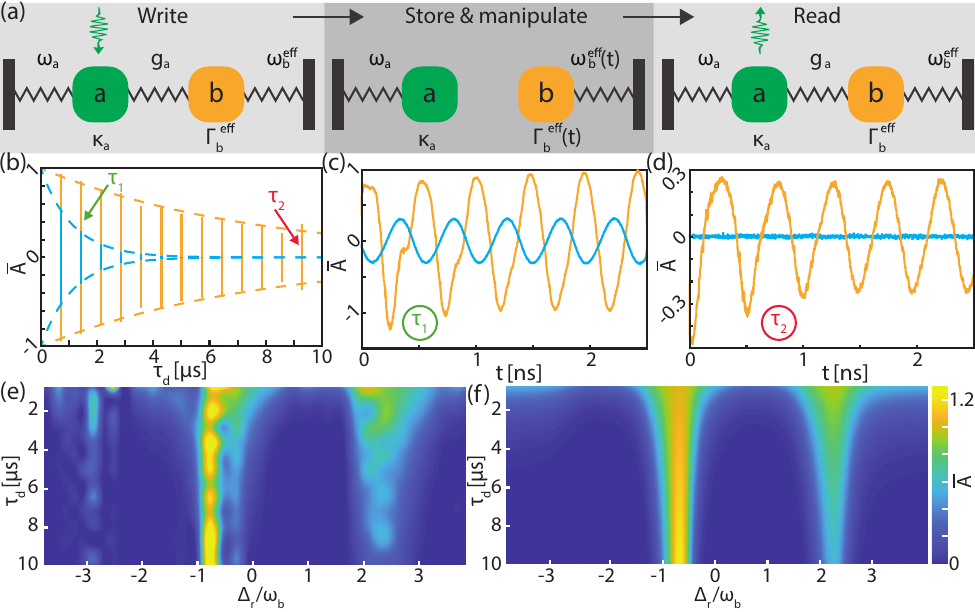}
\caption{{\bf Enhancing the optomechanical pulse storage time}
(a) Outline of the storage protocol where the optomechanical coupling between optical mode $a$ and the mechanical mode facilitates storage of an optical pulse as a mechanical excitation. Mode $r$ is used to manipulate the mechanical damping rate, $\Gamma_{\mathrm{b}}^\mathrm{eff}$(t), and frequency, $\omega_{\mathrm{b}}^\mathrm{eff}$(t) which can be carried out concurrently with the storage process. (b) Normalized read pulse amplitude, $\bar{A}$ for the off-resonant, $\Delta_\mathrm{r} \gg \omega_\mathrm{m}$, case (cyan) and for $\Delta_\mathrm{r} \sim \omega_\mathrm{m}$ (orange) as a function of $\tau_\mathrm{d}$. These are fit to an exponential decay (dashed lines) to extract $\Gamma_\text{b}^\text{eff}$.  (c,d) Zoom-in of of the read data shown in (b) for $\tau_\text{d} = (0.14, 0.93)\, \mu$s, illustrating the enhancement in $\bar{A}$ when $\Delta_\mathrm{r} \sim \omega_\mathrm{m}$.(e) Complete data set from which (b) was taken showing both an enhancement and reduction in storage time as a function of $\Delta_\mathrm{r}$, which is compared to the expected behavior (f).}
\label{fig:2}
\end{figure*}

\subsection{Pulse storage manipulation}

Pioneering experiments with $\Lambda$-type atomic systems have demonstrated that a strong control field can dramatically alter the optical
\dds{response}
of a material,
including rendering otherwise opaque materials transparent \cite{ref:boller1991oei}, enhancing nonlinear processes \cite{ref:jain1995esfpt,ref:hakuta1997sifm}, and slowing the group velocity of a pulse of light \cite{ref:kasapi1995eitdp,ref:hau1999lsr}. Furthermore, by dynamically altering the transparency of a material, a pulse of light may be trapped and deterministically released at a later time \cite{ref:liu2001oco,ref:phillips2001sla,ref:Sprague2014FiberMemory}. Such schemes have been used to store light pulses for over a minute \cite{ref:heinze2013sl1m}, and have been proven \dds{to work at a single-photon level} \cite{ref:chaneliere2005ssp,ref:eisaman2005eit}.

Cavity optomechanical systems have similar capabilities. In OMIT, the interaction between optical mode $a$ and the mechanical resonator  is described by the beamsplitter Hamiltonian $\hat{H}_\text{bs}=\hbar g_\text{a} \left(\hat{a}^\dagger\hat{b} + \hat{a} \hat{b}^\dagger \right)$ when the mode $a$ control field is red detuned by $\omega_\text{b}$ from resonance  \cite{ref:aspelmeyer2014co}. Here $\hat{a}$ $\left(\hat{a}^\dagger\right)$ and $\hat{b}$ $\left(\hat{b}^\dagger\right)$ are annihilation (creation) operators for optical mode $a$ and mechanical resonator mode $b$, respectively.
\dds{This Hamiltonian allows for coherent exchange of excitations  between modes $a$ and $b$.}
By adjusting the control field amplitude, which in turn controls $g_\text{a}$, a field input to $a$ can be coherently and reversibly stored in the mechanical resonator \cite{ref:fiore2011soo, ref:fiore2013ops, ref:palomaki2013CohTrans, ref:walluks2019Memory}.

\peb{ Here we show that 	\dds{an optical field}
stored in the motion of an optomechanical memory can be dynamically modified by varying the resonator's coupling to a reservoir mode.} Our memory protocol is illustrated using mass-on-spring systems in Fig.~\ref{fig:2}(a). During the write stage, the red-detuned mode $a$ control laser couples a weak signal field resonant with mode $\hat{a}$ to mode $\hat{b}$ at rate $g_\mathrm{a}$.
Following the write stage the control laser is removed, decoupling modes $a$ and $b$. Finally, during the read stage, the reservoir control laser is removed and the signal control laser is turned back on, reconverting the stored mechanical signal to the optical domain.
Our scheme deviates from a conventional optomechanical memory\cite{ref:fiore2013ops, ref:palomaki2013CohTrans, ref:walluks2019Memory}  in two ways: not only does coupling to the reservoir  modify the mechanical resonator dynamics
as described above, the reservoir coupling can also be varied temporally. This both modifies the stored information
and the rate at which it dissipates. Finally, during the read stage, the reservoir control laser is removed and the signal control laser is turned back on, reconverting the stored mechanical signal to the optical domain.

Through continuous amplification of the stored mechanical signal  by the reservoir mode, the pulse storage lifetime of the mechanical resonator can be extended. This is in a similar spirit to previous demonstrations of optomechanical amplification in waveguides \cite{ref:stiller2019cra}. \peb{In cavity optomechanics, amplification is achieved by setting $\Delta_\mathrm{r}$ to $+\,\omega_\mathrm{b}$, so that the   reservoir-resonator interaction is governed by the parametric amplifier Hamiltonian $\hat{H}_\text{amp}=\hbar g_\text{r} \left(\hat{r}\hat{b} + \hat{r}^\dagger \hat{b}^\dagger \right)$, where  $\hat{r}$ $\left(\hat{r}^\dagger\right)$ is the annihilation (creation) operator for optical mode $r$. In the experiments $\Delta_\mathrm{r}$ is tuned until $\Gamma_\mathrm{b}^\mathrm{eff}$ nearly vanishes, as in Fig.\ \ref{fig:1}(c). Note that in contrast to Brillouin scattering, the mechanical mode amplitude can grow exponentially during this interaction \cite{ref:aspelmeyer2014co, ref:van2016ubs}.}

To measure the storage time for a given reservoir setting, we varied the delay $\tau_\mathrm{d}$ between the write and read pulses. The measured amplitude output during the read pulse encodes the temporal envelope of the mechanical signal, which decays exponentially at a rate $\Gamma_\mathrm{b}^\mathrm{eff}$. An example of this decay is plotted in Fig.~\ref{fig:2}(b) for two cases: when the reservoir control is optimally detuned for enhanced storage time $(\Delta_\mathrm{r} \approx -\omega_\mathrm{b}$ and $\Gamma_\mathrm{b}^\mathrm{eff} \ll \Gamma_\mathrm{b}$), and when the reservoir is far detuned $(\Delta_\mathrm{r} \gg \omega_\mathrm{b}$ and  $\Gamma_\mathrm{b}^\mathrm{eff} \sim \Gamma_\mathrm{b})$.

Comparing the observed decay rates  for each configuration indicates a $7 \times$ enhancement in 1/$e$ time from  1.1 $\mu$s to 7.7 $\mu$s due to the reservoir coupling. Examples of the signal extracted at two values of $\tau_\mathrm{d}$ are shown on Fig.~\ref{fig:2}(c,d), demonstrating the dramatic increase in the amplitude of the enhanced signal at longer timescales. A full measurement of the readout amplitude decay for varying  $\Delta_\mathrm{r}$ was also acquired, the results of which are plotted in Fig.~\ref{fig:2}(e). This shows qualitative agreement to the theoretical amplitudes obtained from an analytic fit of $\Gamma_\mathrm{b}^\mathrm{eff}(\Delta_\mathrm{r})$ and plotted in Fig.~\ref{fig:2}(f). Note  the doublet nature of the reservoir mode is again evident in the measurement. Differences between theory and experiment are understood to be a consequence of thermal drifts in the resonance frequencies of the modes, as well as wavelength drifts in the readout laser used in the experiment.
\dds{These imperfections lead to a deviation of the readout signal from a sinusoid in Fig.~\ref{fig:2}(c) and will reduce the fidelity of the reamplified memory. They can be mitigated by an active stabilization of $\Delta_r$ but further measurements with shaped input pulses are required to fully characterize their effect.
}

\dds{The optomechanical memory's storage time depends on dephasing governed by $1/\Gamma_\text{b}^\text{eff}$ and on added noise. The latter is created by the optical field, which can amplify thermal phonons and spontaneously generate phonons, in addition to amplifying desirable signal phonons. Ignoring spontaneous processes, given a stored signal with an initial signal-to-noise ratio (SNR$_0$) defined by the ratio of  the signal and thermal phonon populations,  for large SNR$_0$ the thermal and signal phonon populations become equal at  time $t_s \sim 1/\Gamma_\text{b}^\text{eff}\ln\left(\text{SNR}_0\right)$. This storage time exceeds the $1/e$ decay time, as it is defined by the minimum acceptable SNR of the retrieved signal. It ignores the reduction to the initial thermal population from the OMIT write step, which will further extend $t_s$; a more general expression is given together with an analysis of storage time in Supplementary Material, Section IV.} \peb{The maximum SNR$_0$ is set by the device's temperature combined with its maximum achievable oscillation amplitude. From previous measurements an SNR$_0$ of 1000 at room-temperature should be achievable \cite{ref:mitchell2016scd}. \markup{Together with a realistic value} of $\Gamma_\text{b}^\text{eff} = 2\pi \times 1$ kHz, a memory time of $\approx 5$ ms is predicted. Finally, note that the ultimate limit on storage time will be set by generation and amplification of Stokes phonons, which were not considered in this analysis but will become important when operating in the quantum regime.}

\markup{We also emphasize that thermal phonons from the environment preclude storage of quantum states when operating at room temperature, due to the persistent thermal occupation of the memory. Even in the absence of thermal noise, phonons introduced by the reservoir laser, e.g. through spontaneous Stokes scattering, will degrade the device's ability to store quantum states. }
These effects could potentially be mitigated by utilizing phononic shielding\mjm{\cite{ref:chan2011lcn,ref:nguyen2015iod}}, and by further increasing the optomechanical cooperativity through reducing the device's optical loss (which is limited by surface roughness\mjm{\cite{ref:mitchell2018rqo}}), its mechanical dissipation, and its optical mode volume.

\peb{However, in the measurements of decay time presented here, technical noise dominated over amplified mechanical noise. In particular, drifts in control and probe laser detunings required operating at a lower $P_\text{r}$ compared to the cooperativity enhancement measurements in Section I-A,  to ensure that the system did not enter the unstable regime ($\Gamma_\text{b}^\text{eff} < 0$). This lower power explains the  modest observed decease in the decay time compared to the 158$\times$ cooperativity enhancement reported above. By operating more closely to the self-oscillation threshold through stabilization of $\Delta_\text{r}$, reductions in decay time similar to or exceeding the cooperativity enhancement will be possible.  }

\dds{
Another important characteristic of a memory is its time--bandwidth product (TBP).  A conservative measure of this for our system can be obtained from the product of the $1/e$ decay time and the mechanical linewidth. For CW excitation of the reservoir, this product will not change with $\Gamma_\mathrm{b}^\mathrm{eff}$ as any increase in decay time will be accompanied by a reduction in linewidth; it is 1.38 for our device which is typical for OMIT-based memories \cite{ref:fiore2011soo, ref:fiore2013ops} and is an order of magnitude smaller than in Brillouin waveguide memories \cite{ref:stiller2019cra}. }\markup{{However, note that in waveguide systems storage time is often defined by SNR, as discussed above, resulting in a TBP $\propto \ln\left(\text{SNR}_0\right)$ dependent on the initial signal strength.}}
\peb{Increasing the bandwidth can be realized by using multiple mechanical modes or cascaded resonators\cite{ref:chang2011ssl}, which is predicted to provide a maximum bandwidth $\sim \Gamma_b C$. Alternatively, by operating with a reservoir field that is turned on adiabatically during the storage phase, one could combine enhanced storage times and the intrinsic bandwidth of the mechanical resonator.}

\begin{figure}
\centering
\includegraphics[width=\linewidth]{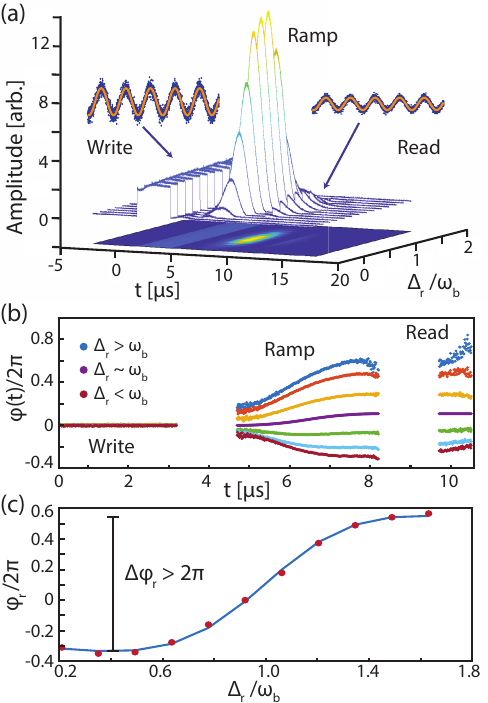}
\caption{{\bf Tuning the phase of the retrieved pulse} (a) Amplitude of the acquired signals vs.\ time and reservoir detuning for the phase manipulation experiment. In this experiment no optical filtering was used, so that the write, ramp, and read pulses may all be detected. (b) Phase of the mechanical oscillator vs.\ time as inferred from the optical output of the device. (c) Final phase as a function of $\Delta_\mathrm{r}$ along with a fit to the data based on the predicted temporally varying optical spring effect associated with the ramp pulse.}
\label{fig:3}
\end{figure}

We next show that the phase of the stored pulse can be controlled by dynamically varying the reservoir mode input. By changing $\omega_\mathrm{b}^\mathrm{eff}$ adiabatically and hence the frequency of the stored pulse, we can complete a trajectory which moves away from and then returns to the original frequency. Over the course of this trajectory, the mechanical oscillator acquires a dynamical phase $\varphi(t_2) = \varphi(t_1) + \int_{t_1}^{t_2} \delta\omega(t) dt$, assuming that we return to the original mechanical frequency. This is analogous to a pendulum whose length is adjusted in time \cite{ref:goldstein2002clm}. In our experiment, we varied the amplitude of the reservoir mode in time using an amplitude electro-optic modulator driven by a symmetric RF ramp pulse (see Supplementary Material) for various $\Delta_\mathrm{r}$ as shown in Fig.~\ref{fig:3}(a). Here the ramp pulse was 3.5 $\mu$s long and was situated 1.5 $\mu$s after the write, and before the read pulse. By fitting the beat note detected at the fiber output for each of the write, ramp, and read pulse segments we can plot the phase as referenced to the well-defined write pulse, as shown in shown in Fig.~\ref{fig:3}(b). Here we have removed the phase shift associated with the spring effect induced by the write pulse, which added a linear slope to the ramp and read pulse segments. This allows us to isolate the shift due to the spring effect associated with the reservoir mode ramp pulse (see Supplementary Material).  From the read pulse segments we extract the phase relative to the write pulse, demonstrating a reservoir controlled phase shift, $\Delta\varphi > 2\pi$ as shown in Fig.~\ref{fig:3}(c).\peb{Further studies of quantum noise generated by the reservoir mode during the frequency tuning process are required to assess the performance of this technique in the quantum domain.}

\section{Discussion}

\peb{ In this work we have demonstrated in-situ control of an optomechanical memory,} bypassing the usual limitations imposed by the intrinsic damping rate and generating controllable phase shifts in the stored information. \peb{The diamond microdisk platform benefits from diamond's superior optical and mechanical properties. Low optical absorption,  high thermal conductivity, and a cavity-based design minimize the necessary control and probe lasers powers, compared to waveguide optomechanics, by creating high intra-cavity light intensities. Low mechanical dissipation ensures that high effective cooperativity and vanishing effective mechanical dissipation can be achieved at the modest power levels used here, resulting in an increased memory lifetime that is primarily limited by the stability of the system not available optical power. Together with a small footprint, this allows for compact, fully integrated optical memories whose multimode nature allows  stored information to be coherently manipulated.}

\peb{In future, performance of the demonstrated optomechanical memory can be improved by cooling and cascading devices to enhance the memory time and memory-bandwidth product. By virtue of operating in the sideband resolved regime, $\Gamma_\mathrm{b}^\mathrm{eff}$ and $\omega_\mathrm{b}^\mathrm{eff}$ for our device are linearly independent of $g_\mathrm{r}$ and $\Delta_\mathrm{r}$ \cite{ref:Xu2016TopTrans}. As a result, it should be possible to incorporate dynamical control of the damping rate,  enabling pulse compression by the realization of a time lens \cite{ref:Kolner1989timeLens, ref:Patera2017TimeLens} (see Supplementary Material) as well as other processing functionality inspired by Brillouin scattering based signal processing \cite{ref:santagiustina2013asp}. Operating the device in the OMIA regime will allow demonstration of narrow spectral filters. Furthermore, spatially-confined and optically-induced mechanical oscillations can be coupled, for example, to spin qubits and electro-magnetic fields, facilitating development of next generation hybrid quantum systems.}


%

\section*{Methods}

\subsection{Fabrication}

The microdisks studied here were fabricated from an optical grade, chemical vapor deposition (CVD) grown $\langle 100 \rangle$-oriented single crystal diamond  substrate supplied by Element Six, according the process outlined in detail in Ref.\cite{ref:mitchell2018rqo}. The polished substrates were first cleaned in boiling piranha, and coated with $\sim$ 400 nm of PECVD Si$_3$N$_4$ as a hard mask. To avoid charging effects during electron beam lithography (EBL), $\sim$ 5 nm of Ti was deposited before the ZEP 520A EBL resist. The developed pattern was transferred to the hard mask via inductively coupled reactive ion etching (ICPRIE) with C$_4$F$_8$/SF$_6$ chemistry. The anisotropic ICPRIE diamond etch was performed using O$_2$, followed by deposition of $\sim$ 250 nm of conformal PECVD Si$_3$N$_4$ as a sidewall protection layer. The bottom of the etch windows were then cleared of Si$_3$N$_4$ using a short ICPRIE etch with C$_4$F$_8$/SF$_6$. This was followed by a zero RF power O$_2$ RIE diamond undercut etch to partially release the devices. The undercutting process was interrupted and an $\sim 100 $ nm layer of SiO$_2$ is deposited via electron beam evaporation to alter the microdisk pedestal profile before finishing the undercut. Lastly, the Si$_3$N$_4$ layer was removed using a wet-etch in 49$\%$ HF, and the devices were cleaned again in boiling piranha.

\subsection{Apparatus}

\begin{figure}
\centering
\includegraphics[width=0.92\linewidth]{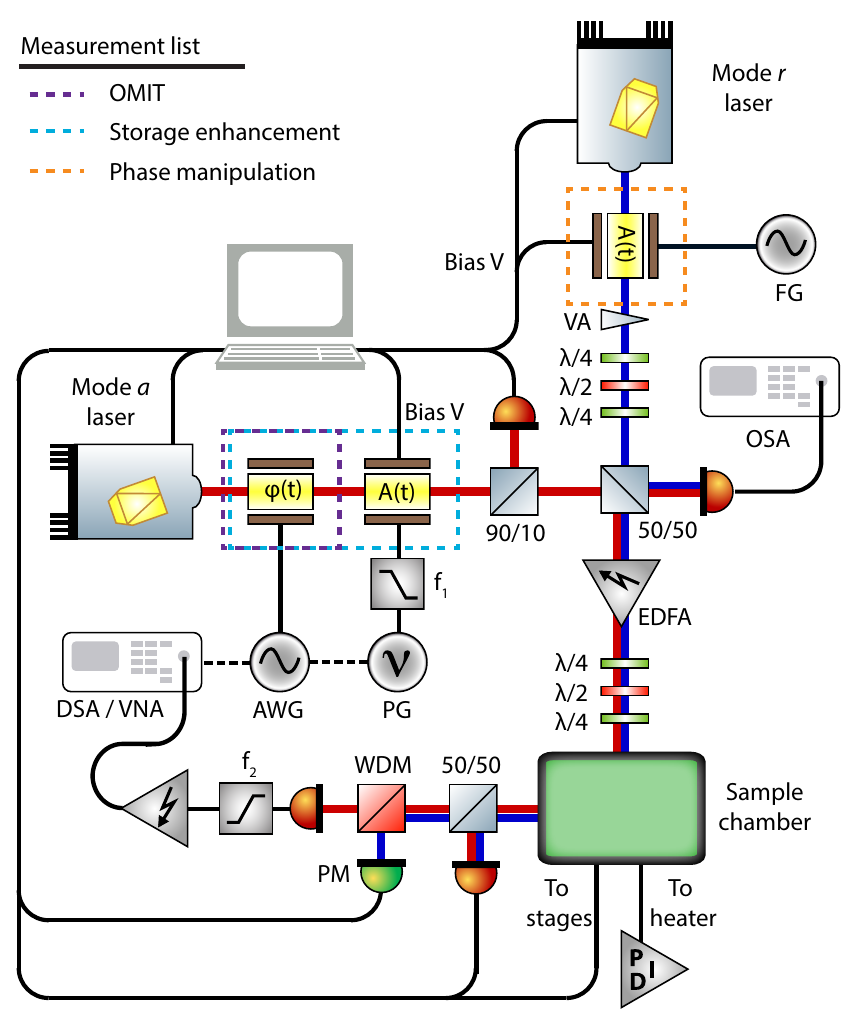}
\caption{{\bf Experimental apparatus} Measurement setup including the components necessary for all measurements described in the main text. Legend indicates the optical components necessary for each measurement presented and the components are discussed in detail in the text.}
\label{fig:supp1}
\end{figure}

The measurement apparatus for the results described in the main text is shown in Fig.\ \ref{fig:supp1}, where the optical components added for each measurement described in the main text are outlined in the legend. Mode $a$ was driven by a tunable diode laser at 1560 nm (Newport TLB-6700) while the reservoir mode, $r$,  was also driven by a tunable diode laser at 1520 nm (Newport TLB-6700) whose output was connected to a variable attenuator (VA: Exfo FVA-3100). For the verification of mutual coherence the mode $a$ laser was passed through a phase electro-optic modulator ($\varphi(t)$: EOSpace PM-5S5-20-PFA-PFA-UV-UL) to generate a weak probe field from the control field, which was swept across the resonance using a vector network analyzer (VNA: Keysight E5063A) allowing the measurement of OMIT. The mode $r$ laser was combined with the mode $a$ laser on a fiber coupled 50/50 beamsplitter (BS: Newport F-CPL-L22355-A) where one output was sent to an erbium doped fiber amplifier (EDFA: Pritel LNHPFA-30-IO) and the other to an optical spectrum analyzer (OSA: Ando AQ6317B) such that the laser wavelengths could be tracked during the experiment. The output of the EDFA was sent to a fiber polarization unit followed by the fiber taper waveguide coupled to the diamond microdisk. The output of the fiber taper was then split on another 50/50 beam-splitter where one output is sent to a power meter (PM: Newport 1936-R) and one to a 1510/1550 nm wavelength division multiplexer (WDM: Montclair MFT-MC-51-30 AFC/AFC-1) to separate the light from modes $a$ and $r$. The output of the WDM was then measured on a high speed photodetector (New Focus 1554-B) whose output is high-pass filtered and electronically amplified before being sent to the VNA for measuring OMIT or a digital serial analyzer (DSA:Tektronix DSA70804B) for the pulse storage measurements.

\begin{figure}
\centering
\includegraphics[width=0.95\linewidth]{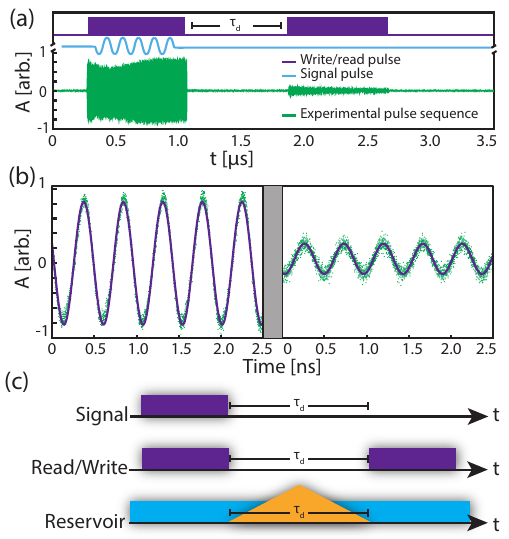}
\caption{{Pulse storage measurement procedure} (a) Timing diagram for optomechanical pulse storage, along with an example of a typical measure signal (purple). (b) Detail of a short section of the write (left panel) and read pulse (right panel). (c) Timing diagram for the storage enhancement and phase manipulation experiments described in the main text.}
\label{fig:supp2}
\end{figure}

In the pulse storage enhancement measurement an amplitude modulator ($A(t)$: EOSpace AZ-0K5-10-PFA-SFA) was added to the mode $a$ laser to generate the optical pulses, while the phase modulator was used to generate the signal to be written. The signal to be written was a sine wave at $\omega_\text{b}$ generated by an arbitrary waveform generator (AWG: Tektronix AWG70002A) which was superimposed on the optical pulses generated by a low pass filtered pulse generator (PG: Stanford Research Systems DG535), triggered by the AWG. Here the reservoir mode control laser was operated in c.w. mode, however, during the phase manipulation measurement an amplitude modulator ($A(t)$: Lucent Technologies 2623CS) is added to the output of the laser, which was driven by a separate function generator to generate the phase manipulation pulses, as shown in Fig.~\ref{fig:supp2}. A thermoelectric heater/cooler was also placed under the sample and controlled with a PID for thermal stability during the measurement.

\subsection{Mode characterization}

\begin{figure}[ht]
\centering
\includegraphics[width=0.95\linewidth]{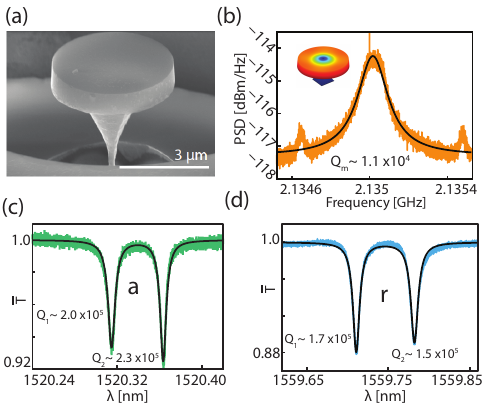}
\caption{{\bf Characterization of optical and mechanical modes} (a) Scanning electron micrograph of the diamond microdisk used in this work. (b) Measured power spectral density of radial breathing mode at room temperature and pressure (inset: COMSOL simulated displacement profile). (c,d) Fiber taper transmission scans for both optical modes used in this work, revealing the doublet nature of the modes. Intrinsic optical quality factors are labelled.}
\label{fig:supp_modes}
\end{figure}

 The power spectral density for the radial breathing mode was measured using a high speed photodetector and real time spectrum analyzer (Tektronix RSA5106A), and is shown in Fig.\ \ref{fig:supp_modes}(b), which was carried out at low input power $(P_\mathrm{in}\sim50\mu \mathrm{W})$ to avoid optomechanical backaction. By fitting the power spectral density to a Lorentzian we extract a mechanical quality factor, $Q_\mathrm{m} \sim 1.1\times10^4$, at room temperature and pressure. Optical transmission scans are shown in Fig.\ \ref{fig:supp_modes}(c,d), with intrinsic quality factors labelled for each of the doublet modes. The per-photon optomechanical coupling rates were measured in a separate experiment, yielding $g_\mathrm{r}/\alpha_\mathrm{r},~ g_\mathrm{a}/\alpha_\mathrm{a} \sim 2\pi\times 25 \mathrm{kHz}$, where $\alpha_\mathrm{r}$ and $ \alpha_\mathrm{a}$ are the strong control laser amplitudes for mode $r$, and $a$, respectively (see Supplementary Material).

\vspace{3mm}

 In the limit that surface scattering effects are smaller than all other optical loss mechanisms ($g_\text{ss}\ll\kappa$), microdisks will possess degenerate clockwise and anticlockwise propagating modes, with negligible coupling between them. However, when the surface scattering approaches or exceeds the optical linewidth ($g_\text{ss}\ge\kappa$), the clockwise and counter clockwise modes couple and will form pairs of modes known as optical doublets. These are simply symmetric and antisymmetric combinations of the travelling wave modes. The orthogonality of the doublet modes allow us to calculate the overall mechanical frequency shift, or damping rate induced by a strong control laser by taking the sum of the contributions from the symmetric and antisymmetric mode.

\section*{\textbf{Data Availability}}
\noindent The datasets generated during and/or analyzed during the current study are available from the corresponding author on reasonable request.

\section*{\textbf{Author contributions}}

\noindent D.P.L. and M.M. fabricated the device, performed the experiments, and analyzed the data. D.P.L. and D.S. developed the theory.  P.E.B. supervised the project. All authors contributed to writing the manuscript.

\section*{\textbf{Acknowledgments}}

\noindent The authors would like to thank Daniel Oblak and the QCloudLab for their generosity in lending us some of the components required to obtain these results. This work was supported by the National Research Council, Canadian Foundation for Innovation, Alberta Innovates, and the Natural Sciences and Engineering Research Council of Canada.

\section*{\textbf{Competing interests}}

\noindent The authors declare no competing interests.

\clearpage

\newpage

\setcounter{equation}{0}
\setcounter{figure}{0}
\setcounter{section}{0}
\setcounter{subsection}{0}
\setcounter{table}{0}
\setcounter{page}{1}
\makeatletter
\renewcommand{\theequation}{S\arabic{equation}}
\renewcommand{\thetable}{S\arabic{table}}
\renewcommand{\thefigure}{S\arabic{figure}}
\renewcommand{ \citenumfont}[1]{S#1}
\renewcommand{\bibnumfmt}[1]{[S#1]}

\onecolumngrid
\section*{Supplementary Information}

\section{Tuning the system dynamics}

The results of this work are enabled by dynamic manipulation of a mechanical oscillator through engineering the degrees of freedom it is coupled to. The manifestation of this coupling is optomechanical damping, and the optomechanical spring shift, which have been extensively studied in the past \cite{ref:supp_marquardt2007sbcool,ref:supp_Wilson-Rae2007GSCool}. These previous approaches leveraged frequency domain calculations and a Kubo formula to calculate damping rates, and the minimum phonon occupation. The frequency shift was then calculated through the Kramers Kronig relations. In our approach, we directly solve expressions in the time domain in the form of retarded Green's functions, and then adiabatically eliminate the fast dynamics of the cavity.

In the absence of optomechanical coupling, the mechanical mode is modelled as a damped harmonic oscillator, with annihilation operator $\hat{b}$, intrinsic frequency $\omega_\text{b}$, and intrinsic damping rate $\Gamma_\text{b}$. The damping is a consequence of coupling to the environment, and according to the fluctuation-dissipation theorem this damping will be accompanied by a dissipation term. In the input-output formalism, we write
\begin{equation}\label{eq:mechEnvironment}
\dot{\hat{b}} = -\left(i\omega_\text{b}+ \frac{\Gamma_\text{b}}{2}\right)\hat{b} + \sqrt{\Gamma_\text{b}}\hat{e}_\text{in},
\end{equation}
where $\hat{e}_\text{in}$ is the input from the environment. $\omega_\text{b}$ and $\Gamma_\text{b}$ are intrinsic properties of the device, however cavity optomechanics offers a way to manipulate the mechanical parameters of the cavity by optical means \cite{ref:supp_aspelmeyer2014co}. As we will show here, this can be viewed as coupling the mechanical mode to an optical reservoir mode, $\hat{r}$. Not only is this interaction easily controllable, but the reservoir can be arranged to have negligible thermal occupation through use of optical laser light.

We consider a reservoir mode with frequency $\omega_\text{r}$ connected to an input port at a rate $\kappa_\text{r}^\text{ex}$ with a total decay rate $\kappa_\text{r}$. This is dissipatively coupled to the mechanical mode with a strength $g_\text{r}$. In the presence of a strong control laser with amplitude $\alpha_\text{r}$, the optomechanical interaction can be linearized, and we can write expressions for the cavity fluctuation operator $\hat{r}$, which is coupled the mechanics at a rate $g_\text{r}=G_\text{r} \alpha_\text{r} x_\text{o}$, where $G_\text{r} = \frac{d\omega}{d x}$ is the shift in cavity frequency due mechanical displacement, and $x_\text{o}$ are the zero point fluctuations of the mechanics \cite{ref:supp_aspelmeyer2014co}. This leads to the coupled equations of motion
\begin{equation}\label{eq:EOMMatrix}
\begin{bmatrix}
\left(\frac{d}{dt} + \frac{1}{\chi_\text{b}}\right) & 0 & ig_\text{r} & ig_\text{r} \\
0 & \left(\frac{d}{dt} + \frac{1}{\chi_{\text{b}^\dagger}}\right) & - ig_\text{r} & - ig_\text{r} \\
ig_\text{r} & ig_\text{r} & \left(\frac{d}{dt} + \frac{1}{\chi_\text{r}}\right) & 0 \\
-ig_\text{r} & -ig_\text{r} & 0 & \left(\frac{d}{dt} + \frac{1}{\chi_{\text{r}^\dagger}}\right)
\end{bmatrix}
\begin{bmatrix}
\hat{b} \\
\hat{b}^\dagger\\
\hat{r} \\
\hat{r}^\dagger
\end{bmatrix}
=
\begin{bmatrix}
\sqrt{\Gamma_\text{b}}\hat{e}_\text{in} \\
\sqrt{\Gamma_\text{b}}\hat{e}^\dagger_\text{in}\\
\sqrt{\kappa_\text{r}^\text{ex}}\hat{r}_\text{in} \\
\sqrt{\kappa_\text{r}^\text{ex}}\hat{r}^\dagger_\text{in}
\end{bmatrix},
\end{equation}
where we define the relevant response functions as $\chi_\text{b}^{-1}(\omega) = \Gamma_\text{b}/2-i(\omega-\omega_\text{b})$, $\chi_{\text{b}^\dagger}^{-1}(\omega) = \Gamma_\text{b}/2-i(\omega+\omega_\text{b})$, $\chi_\text{r}^{-1}(\omega) = \kappa_\text{r}/2-i(\omega+\Delta_\text{r})$, and $\chi_{\text{r}^\dagger}^{-1}(\omega) = \kappa_\text{r}/2-i(\omega-\Delta_\text{r})$. The input modes at time t, are given in terms of the time $t_0$ in the far past as \cite{ref:supp_gardiner1991qn}
\begin{align}
\hat{e}_\text{in}(t) &=\ \frac{1}{\sqrt{2\pi}} \int e^{-i\omega(t-t_0)}E_0(\omega) d\omega, \\
\hat{r}_\text{in}(t) &=\ \frac{1}{\sqrt{2\pi}} \int e^{-i\omega(t-t_0)}R_0(\omega) d\omega,
\end{align}
where $E_0$ and $R_0$ are the state of the input modes at time $t_0$. For the sake of simplicity, in what follows, we will assume $\kappa_\text{r}=\kappa_\text{r}^\text{ex}$.

Using Eq.\,\ref{eq:EOMMatrix}, we can solve for the reservoir dynamics as
\begin{align}
\hat{r}(t) &=\ \hat{r}_0e^{-(t-t_0)/\chi_\text{r}} + \int_{t_0}^t e^{-(t-t')/\chi_\text{r}} \left( \sqrt{\kappa_\text{r}} \hat{r}_\text{in} + iG_\text{r}\hat{b}+ iG_\text{r} \hat{b}^\dagger\right) dt', \\
\hat{r}^\dagger(t) &=\ \hat{r}_0^\dagger e^{-(t-t_0)/\chi_{\text{r}^\dagger}} + \int_{t_0}^t e^{-(t-t')/\chi_{\text{r}^\dagger}} \left(\sqrt{\kappa_\text{r}} \hat{r}_\text{in}^\dagger -iG_\text{r}\hat{b}- iG_\text{r} \hat{b}^\dagger \right) dt'.
\end{align}
It is interesting to note the role of the optical cavity as a filter. The exponential terms are in the form of a retarded Green's function, and specify a sensitivity to frequencies near $\pm\Delta_\text{r}$ to a history on the timescale $1/\kappa_\text{r}$.

Inserting this into the expression for the mechanics given in Eq.\,\ref{eq:EOMMatrix}, we find an equation of motion for the mechanics under the influence of both the environment and the reservoir
\begin{align}\label{eq:bEOMUnrefined}
\left(\frac{d}{dt} + i\omega_b\right)\hat{b} =&\
-\frac{\Gamma_\text{b}}{2} \hat{b}
- |g_\text{b}|^2 \int_{t_0}^t \left(e^{-(t-t')/\chi_\text{r}}\hat{b}(t') - e^{-(t-t')/\chi_{\text{r}^\dagger}} \hat{b}(t') \right)dt' \nonumber\\
& +\sqrt{\Gamma_\text{b}}\hat{e}_\text{in}
+ iG_\text{b}\sqrt{\kappa_\text{r}} \int_{t_0}^t \left(e^{-(t-t')/\chi_\text{r}} \hat{r}_\text{in}(t') + e^{-(t-t')/\chi_{\text{r}^\dagger}} \hat{r}_\text{in}^\dagger(t')\right)dt',
\end{align}
where in the above we used the fact the $t-t_0\gg 1/\kappa_\text{r}$ and applied the rotating wave approximation. The right side of the equation can be interpreted as the sum of damping and dissipation terms due to coupling to the environment, and damping and dissipation terms due to coupling to the reservoir. With the assumption that $ \kappa_\text{r} \gg \Gamma_\text{b}$, we can make further simplifications. First we note that the integral associated with the dissipation term becomes
\begin{align}\label{eq:dissipationUnrefined}
\int_{t_0}^t \left(e^{-(t-t')/\chi_\text{r}} + e^{-(t-t')/\chi_{\text{r}^\dagger}} \right) \hat{b}(t') dt'
\approx&\ \int_{t_0}^t \left(e^{-(t-t')/\chi_\text{r}} - e^{-(t-t')/\chi_{\text{r}^\dagger}} \right) e^{i\omega_b(t-t')}\hat{b}(t) dt' \nonumber\\
=&\ \left(\chi_\text{r}(\omega_b) - \chi_{\text{r}^\dagger}(\omega_b)\right) \hat{b}(t).
\end{align}
Next we simplify the fluctuation term as
\begin{align}\label{eq:fluctuationUnrefined}
&\int_{t_0}^t \left(e^{-(t-t')/\chi_\text{r}} \hat{r}_\text{in}(t') + e^{-(t-t')/\chi_{\text{r}^\dagger}} \hat{r}_\text{in}^\dagger(t')\right)dt' \nonumber\\
&=\frac{1}{\sqrt{2\pi}} \int \int_{t_0}^t \left(e^{\left(\chi_\text{r}^{-1}-i\omega\right)t'} e^{-\chi_\text{r}^{-1} t+i\omega t_0} \hat{R}_0(\omega)+ e^{\left(\chi_{\text{r}^\dagger}^{-1}+i\omega\right)t'} e^{-\chi_{\text{r}^\dagger}^{-1}t-i\omega t_0} \hat{R}_0^\dagger(\omega) \right) dt' d\omega \nonumber\\
&=\frac{1}{\sqrt{2\pi}}\int \left(\frac{e^{(\chi_\text{r}^{-1}-i\omega)t}-e^{(\chi_\text{r}^{-1}-i\omega)t_0}}{\chi_\text{r}^{-1}-i\omega} e^{-\chi_\text{r}^{-1} t+i\omega t_0} \hat{B}_0(\omega)+
\frac{e^{(\chi_{\text{r}^\dagger}^{-1}+i\omega)t}-e^{(\chi_{\text{r}^\dagger}^{-1}+i\omega)t_0}}{\chi_{\text{r}^\dagger}^{-1}+i\omega} e^{-\chi_{\text{r}^\dagger}^{-1}t-i\omega t_0} \hat{R}_0^\dagger(\omega) \right) d\omega \nonumber\\
&\approx \frac{1}{\sqrt{2\pi}} \int \left( e^{-i\omega(t-t_0)} \chi_\text{r}(\omega) \hat{R}_0(\omega) + e^{i\omega(t-t_0)} \chi_{\text{r}^\dagger}(\omega) \hat{R}_0^\dagger(\omega) \right) d\omega \nonumber\\
&\approx \frac{1}{\sqrt{2\pi}} \int \left( e^{-i\omega(t-t_0)} \chi_\text{b}(\omega_\text{b}) \hat{R}_0(\omega) + e^{i\omega(t-t_0)} \chi_{\text{r}^\dagger}(\omega_\text{b}) \hat{R}_0^\dagger(\omega) \right) d\omega \nonumber\\
&= \chi_\text{r}(\omega_\text{b})\hat{r}_\text{in} + \chi_{\text{r}^\dagger}(\omega_b)\hat{r}_\text{in}^\dagger.
\end{align}
where once again we used the assumption that $ \kappa_\text{r} \gg \Gamma_\text{b}$ to simplify. Combining Eqs.\,\ref{eq:bEOMUnrefined}--\ref{eq:fluctuationUnrefined} we arrive at the solution
\begin{align}
\left(\frac{d}{dt} + i\omega_\text{b} \right)\hat{b}=& -\left(\frac{\Gamma_\text{b}}{2} + |g_\text{r}|^2\chi_\text{r}(\omega_\text{b}) - |g_\text{r}|^2\chi_{\text{r}^\dagger}(\omega_\text{b}) \right)\hat{b} \nonumber\\
&+\sqrt{\Gamma_\text{b}}\hat{e}_\text{in} + iG_\text{r} \sqrt{\kappa_\text{r}}\chi_\text{r}(\omega_\text{b})\hat{r}_\text{in} + iG_\text{r} \sqrt{\kappa_\text{r}}\chi_{\text{r}^\dagger}(\omega_\text{b})\hat{r}_\text{in}^\dagger.
\end{align}
This can be rearranged to the simple expression reminiscent of Eq.\,\ref{eq:mechEnvironment}
\begin{equation}\label{eq:supp_mechReservoir}
\dot{\hat{b}} = -\left(i\omega_\text{b}^\text{eff} + \frac{\Gamma_\text{b}^\text{eff}}{2} \right)\hat{b}+\sqrt{\Gamma_\text{b}}\hat{e}_\text{in}+g_\text{r} \sqrt{\kappa_\text{r}}\chi_\text{r}(\omega_\text{b})\hat{r}_\text{in} + g_\text{r} \sqrt{\kappa_\text{r}}\chi_{\text{r}^\dagger}(\omega_\text{b})\hat{r}_\text{in}^\dagger.
\end{equation}
In the above we have absorbed a factor of $i$ into the definition of $R_0$ and $R_0^\dagger$, and defined effective frequency and damping terms
\begin{align}
\omega_\text{b}^\text{eff} =&\ \omega_\text{b} + \omega_\text{r}^\text{opt} =\ \omega_\text{b} +
|g_\text{r}|^2\left(\frac{\omega_\text{b}+\Delta}{\kappa_\text{r}^2/4+\left(\omega_\text{b} + \Delta_\text{r} \right)^2} + \frac{\omega_\text{b}-\Delta}{\kappa_\text{r}^2/4+\left(\omega_\text{b} - \Delta_\text{r} \right)^2} \right), \\
\Gamma_\text{b}^\text{eff} =&\ \Gamma_\text{b} + \Gamma_\text{r}^\text{opt}=\ \Gamma_\text{b} +
|g_\text{r}|^2\left(\frac{\kappa_\text{r}}{\kappa_\text{r}^2/4+\left(\omega_\text{b} + \Delta_\text{r} \right)^2} - \frac{\kappa_\text{r}}{\kappa_\text{r}^2/4+\left(\omega_\text{b} - \Delta_\text{r} \right)^2} \right).
\end{align}

Comparing Eq.\,\ref{eq:mechEnvironment} and Eq.\,\ref{eq:supp_mechReservoir}, we see that coupling the reservoir mode induces both fluctuation and dissipation. By varying the strength or detuning of the control laser, the coupling to the reservoir is modified. In the sideband resolved regime $(\omega_\text{b} \gg \kappa)$ we note two special cases. For $\Delta_\text{r} = -\omega_\text{b}$ the effective interaction Hamiltonian is $H_\text{eff} = -g_\text{r} \left(\hat{b}^\dagger\hat{r} + \hat{b}\hat{r}^\dagger\right)$, and the mechanics has the equation of motion
\begin{equation}
\dot{\hat{b}} = -\left(i\omega_\text{b} + \frac{\Gamma_\text{b}}{2}+ \frac{\Gamma_\text{r}^\text{opt}}{2} \right)\hat{b}
+\sqrt{\Gamma_\text{b}}\hat{e}_\text{in} +\sqrt{\Gamma_\text{r}^\text{opt}}\hat{r}_\text{in}^\dagger.
\end{equation}
On the other hand, for $\Delta_\text{r} = \omega_\text{b}$ the interaction Hamiltonian takes the form $H_\text{eff} = -g_\text{r} \left(\hat{b}\hat{r} + \hat{b}^\dagger\hat{r}^\dagger\right)$, and the equation of motion is
\begin{equation}
\dot{\hat{b}} = -\left(i\omega_\text{b} + \frac{\Gamma_\text{b}}{2} - \frac{\Gamma_\text{r}^\text{opt}}{2} \right)\hat{b}
+\sqrt{\Gamma_\text{b}}\hat{e}_\text{in} + \sqrt{\Gamma_\text{b}}\hat{r}_\text{in}.
\end{equation}

\section{Enhanced OMIT}

The amplitude in cavity $a$, as a function of probe-control field detuning, $\delta_\mathrm{a}$, under the influence of the reservoir mode may be expressed as,
\begin{equation}
a(\delta_\mathrm{a}) = -\frac{\sqrt{\kappa_\text{ex}}\hat{a}_\text{in}(\omega)}{i(-\omega_\text{b}+\delta_\mathrm{a})-\kappa_\text{a}/2
-\frac{|g_\text{a}|^2}{i(\omega_\text{b}+\omega_\text{r}^\text{opt}-\delta_\mathrm{a})+(\Gamma_\text{b}+\Gamma_\text{r}^\text{opt})/2}}.
\end{equation}

\noindent From inspection with the OMIT lineshape for a conventional optomechanical system \cite{ref:supp_aspelmeyer2014co}, when our probe is on-resonance, such that $\delta_\mathrm{a}=\omega_\text{b}$, we can write our effective cooperativity as:
\begin{equation}
C_\text{a}^\text{eff}=C_\text{a}\frac{\Gamma_\text{b}}{\Gamma_\text{b}+\Gamma^\text{opt}_\text{r}}=C_\text{a}\frac{\Gamma_\text{b}}{\Gamma_\text{b}^\text{eff}},
\end{equation}
\noindent where $C_\text{a}=4|g_\text{a}|^2/\Gamma_\text{b} \kappa_\text{a}$ is the cooperativity of the device in absence of the reservoir.

\begin{figure}
\centering
\includegraphics[width=0.85\linewidth]{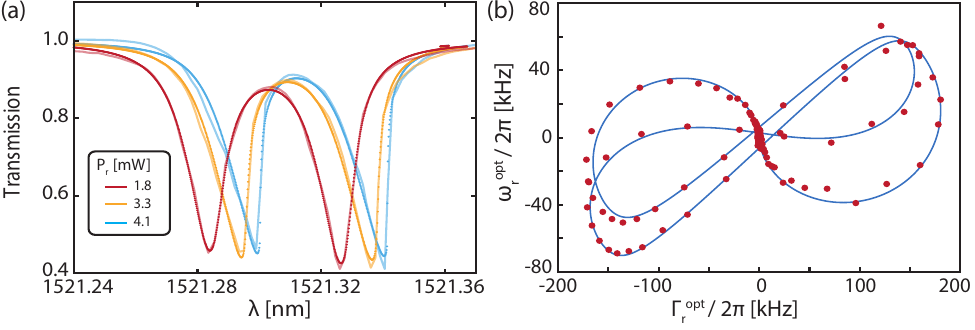}
\caption{
	(a) Optical transmission as a function of laser wavelength of the reservoir mode for increasing input power illustrating the relatively small thermo-optic shift. The $P_\mathrm{r} = 1.8 ~\mathrm{mW}$ curve corresponds to the data presented in the main text. The solid line are fits to the data taking into account thermo-optic effects in the cavity. (b) Optomechanical damping and spring effect due to mode $r$ plotted as functions of each other, corresponding to the data shown in the main text. Here the separation of each trajectory is due to a difference in the resonance contrast of each doublet mode.}
\label{fig:supp_enhanced_omit}
\end{figure}

\subsection{Group delay}

\begin{figure*}
\centering
\includegraphics[width=\linewidth]{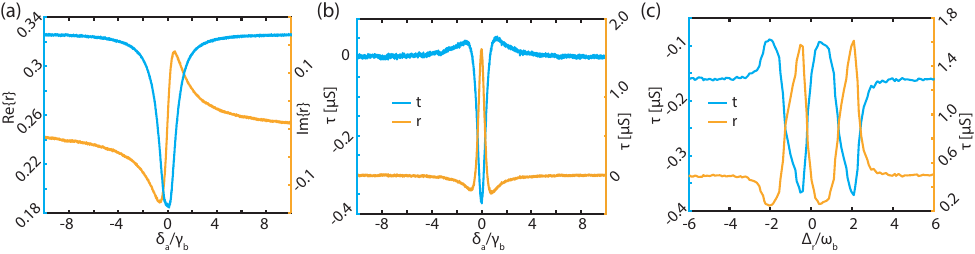}
\caption{
	(a) Real and imaginary parts of the OMIT scan corresponding to the maximum $C_\mathrm{eff}=83$, where $\delta_\mathrm{a}$ is the probe field detuning. (b) Extracted group delay as a function $\delta_\mathrm{a}$, for fixed control laser detuning, $\Delta_\mathrm{a} = \omega_\mathrm{b}$. (c) Extracted group delay as a function of the reservoir mode detuning, $\Delta_\mathrm{r}$, for fixed probe laser detuning, $\delta_\mathrm{a}=\omega_\mathrm{b}$.}
\label{fig:supp_group_delay}
\end{figure*}

The group delay imparted on the pulse in transmission and reflection can be calculated about a central signal frequency, $\omega_\mathrm{s}$ with the spectrum confined to a small window $( < \Gamma^\mathrm{eff}_\mathrm{b})$ following Safavi-Naeini et al. \cite{ref:supp_safavi2011eit} by computing

\begin{equation}
\tau^{(T)} = \mathcal{R}\left\{ \frac{-i}{t(\omega_\mathrm{s})}\frac{dt}{d\omega} \right\},
\end{equation}

\noindent and
\begin{equation}
\tau^{(R)} = \mathcal{R}\left\{\frac{-i}{r(\omega_\mathrm{s})}\frac{dt}{d\omega}\right\},
\end{equation}

\noindent for the transmission and reflection group delay, respectively. These quantities are shown in Fig.~\ref{fig:supp_group_delay} for both fixed and variable reservoir laser detuning.

\section{Cooling and Heating}

As a test of the reservoir engineering expressions, and as a step towards calculating the thermal occupations required for the memory calculations, we calculate full expressions for optomechanical heating and cooling here. Ignoring initial transients, the formal solution of Eq.\,\ref{eq:supp_mechReservoir} is
\begin{equation}
\hat{b}(t) = \int_{t_0}^t e^{-\left(i\omega_\text{b}^\text{eff}+\frac{\Gamma_\text{b}^\text{eff}}{2} \right)\left(t-\tau\right)} \left(\sqrt{\Gamma_\text{b}}\hat{e}_\text{in}+g_\text{r} \sqrt{\kappa_\text{r}}\chi_\text{r}(\omega_\text{b})\hat{r}_\text{in} + g_\text{r} \sqrt{\kappa_\text{r}}\chi_{\text{r}^\dagger}(\omega_\text{b}) \right) d\tau.
\end{equation}
We quantify the thermal statistics of the reservoir and environment with the correlators
\begin{align}
\langle \hat{r}_\text{in}^\dagger(t) \hat{r}_\text{in}(t') \rangle &=\, n_\text{r}^\text{th}\delta(t-t'), \\
\langle \hat{r}_\text{in}(t) \hat{r}_\text{in}^\dagger(t') \rangle &=\, (n_\text{r}^\text{th}+1)\delta(t-t'), \\
\langle \hat{e}_\text{in}^\dagger(t) \hat{e}_\text{in}(t') \rangle &=\, n_\text{e}^\text{th}\delta(t-t'), \\
\langle \hat{e}_\text{in}(t) \hat{e}_\text{in}^\dagger(t') \rangle &=\, (n_\text{e}^\text{th}+1)\delta(t-t')
\end{align}
where $n_\text{r}^\text{th}$ is the number of thermal photons occupying the reservoir, and $n_\text{e}^\text{th}$ is the number of thermal phonons occupying the environment. Using these expressions, we can calculate the thermal occupancy of the cavity as
\begin{align}
\langle \hat{b}^\dagger(t) \hat{b}(t) \rangle =&\ \int_{t_0}^t \int_{t_0}^{t} e^{-\left(-i\omega_\text{b}^\text{eff}+\frac{\Gamma_\text{b}^\text{eff}}{2} \right)\left(t-\tau\right)-\left(i\omega_\text{b}^\text{eff}+\frac{\Gamma_\text{b}^\text{eff}}{2} \right)\left(t-\tau'\right)} \\
&\ \left(\sqrt{\Gamma_\text{b}}\hat{e}_\text{in}(\tau)+g_\text{r} \sqrt{\kappa_\text{r}}\chi_\text{r}(\omega_\text{b})\hat{r}_\text{in}(\tau) + g_\text{r} \sqrt{\kappa_\text{r}}\chi_{\text{r}^\dagger}\hat{r}_\text{in}^\dagger(\tau) \right) \times \nonumber \\
&\ \left(\sqrt{\Gamma_\text{b}}\hat{e}_\text{in}(\tau')+g_\text{r} \sqrt{\kappa_\text{r}}\chi_\text{r}(\omega_\text{b})\hat{r}_\text{in}(\tau') + g_\text{r} \sqrt{\kappa_\text{r}}\chi_{\text{r}^\dagger}\hat{r}_\text{in}^\dagger(\tau') \right) d\tau d\tau' \nonumber\\
=&\ \int_{t_0}^te^{\Gamma_\text{b}^\text{eff}(t-\tau)}\left( \Gamma_\text{b} n_\text{th,b}+\kappa_\text{r}|g_\text{r}\chi_\text{r}(\omega_\text{b})|^2n_\text{th,b} + \kappa_{\text{r}^\dagger}|g_\text{r}\chi_{\text{r}^\dagger}(\omega_\text{b})|^2(n_\text{th,b}+1)\right)d\tau \nonumber \\
=&\ \frac{ \Gamma_\text{b} n_\text{th,b}+\kappa_\text{r}|g_\text{r}\chi_\text{r}(\omega_\text{b})|^2n_\text{th,b} + \kappa_{\text{r}^\dagger}|g_\text{r}\chi_{\text{r}^\dagger}(\omega_\text{b})|^2(n_\text{th,b}+1)}{\Gamma_\text{b}^\text{eff}}
\end{align}

In the experiment considered in this work, our reservoir does not have thermal occupation. Setting $n_\text{r}^\text{th}=0$ we recover the usual limit of optomechanical cooling
\begin{equation}
\braket{\hat{n}} = \frac{\Gamma_\text{b}n_\text{e}^\text{th}+\Gamma_\text{opt}^\text{r}n^\text{min}}{\Gamma_\text{b}+\Gamma_\text{opt}^\text{r}}
\end{equation}
where, $n^\text{min} = |g_\text{r}|^2\kappa\chi_{\text{r}^\dagger}(\omega_\text{b})/\Gamma_\text{r}^\text{opt}$.

\section{Storage Enhancement}

\begin{figure*}
\centering
\includegraphics[width=\linewidth]{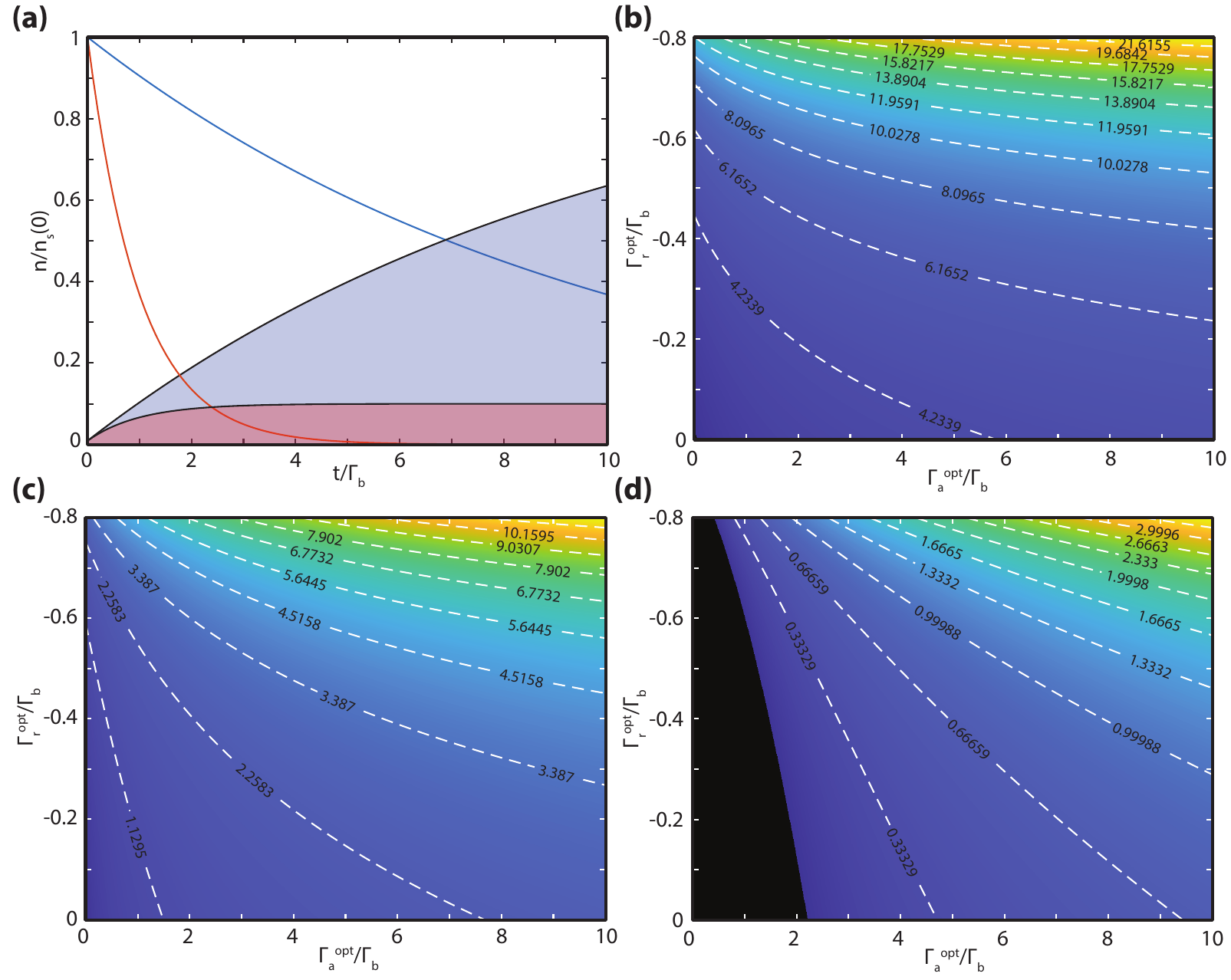}
\caption{
	(a) Signal phonons (solid lines) and thermal phonons (shaded curves) plotted for $\Gamma_\text{r}^\text{opt}/\Gamma_\text{b}=0$ (red) and $\Gamma_\text{r}^\text{opt}/\Gamma_\text{b}=-0.9$ (blue). The initial signal to noise ratio $n_\text{s}(0)/n_\text{th}(0)=10$, and $\Gamma_\text{a}^\text{opt}/\Gamma_\text{b}=10$. (b-d) $t_s \Gamma_\text{b}$ vs. optomechanical damping rates for initial signal to noise ratios $n_\text{s}(0)/n_\text{th}(0)=\{10, 1, 0.1\}$.}
\label{fig:s2}
\end{figure*}

Solving the equations of motion explicitly, we can divide the phonon population in the cavity during the storage time into signal phonons, which are proportional to $\hat{a}_\text{in}$, and undesired thermal phonons, which are a consequence of $\hat{e}_\text{in}$. These each evolve as,
\begin{align}
\langle \hat{b}_\text{s}^\dagger(t)\hat{b}_\text{s}(t) \rangle & = \langle \hat{b_\text{s}}^\dagger(0)\hat{b_\text{s}}(0) \rangle e^{-\Gamma_\text{b}^\text{eff}t} \\
\langle \hat{b}_\text{th}^\dagger(t)\hat{b}_\text{th}(t) \rangle & = n_\text{e}^\text{th}\Gamma_\text{b}
\left(\frac{e^{-\Gamma_\text{b}^\text{eff}t}}{\Gamma_\text{b}+\Gamma_\text{a}^\text{opt}} + \frac{1 - e^{-\Gamma_\text{b}^\text{eff}t}}{\Gamma_\text{b}+\Gamma_\text{r}^\text{opt}}\right).
\end{align}
Here the presence of $\Gamma_\text{a}^\text{opt}$ terms are due to optomechanical cooling by the OMIT control laser during the write step on the initial thermal population of the resonator mode.
Examples of the competing growth and decay of noise and signal phonons is shown in Fig.\ \ref{fig:s2}(a). Defining the storage time as the moment the signal level decays to the level of the thermal phonons, we find,
\begin{equation}\label{eqn:storage_time}
t_s = \frac{1}{\Gamma_\text{b}+\Gamma_\text{r}^\text{opt}}\ln\left( \frac{n_\text{s}(0)}{n_\text{th}(0)} +
\frac{\Gamma_\text{a}^\text{opt}-\Gamma_\text{r}^\text{opt}}{\Gamma_\text{b}+\Gamma_\text{a}^\text{opt}}\right).
\end{equation}
This expression is used to generate the plots in Fig.\ \ref{fig:s2}(b-d), which analyze the performance of the memory as a function of optomechanical damping of modes $a$ and $r$. Note that quantum optical noise, e.g. Stokes scattering, is not included in this analysis.

\section{Phase shifting}

Reservoir engineering also allows us to dynamically change the frequency of the mechanical mode. If the frequency is changed over a time interval $\delta t$, the change in phase may be expressed as,
\begin{equation}
\delta\phi = \int_0^{\delta t}\left(\omega_b(t)-\omega_b(0)\right) dt
\end{equation}
For simplicity, we assume we change our mechanical frequency as a ramp function, with maximum frequency shift $\delta_\text{b}$. Under the adibaticity requirement $1\gg\frac{\delta \omega_\text{b}}{\omega_\text{b}}$.
This yields the simple expression for the phase shift,
\begin{equation}
\delta \phi = \frac{\delta \omega_\text{b} \delta t}{2}.
\end{equation}
In the phase shifting experiment in the main text, we operate with the reservoir laser detuning $\Delta_\text{r} \approx -\omega_\text{b}$, so we may approximate the frequency shift as,
\begin{align}
\delta \omega_\text{b} &\approx \frac{|g_r|^2(\Delta_\text{r}-\omega_\text{b})}{(\Delta_\text{r}-\omega_\text{b})^2+\kappa_\text{r}^2/4}.
\end{align}

\section{Time lens}
The reservoir mode also permits the mechanical damping rate to be dynamically adjusted. For example, at $\Delta_\text{r} \approx -\omega_\text{b}$, the damping is approximately,
\begin{align}
\Gamma_\text{r}^\text{opt} &\approx \frac{-\kappa_\text{r}|g_\text{r}|^2/2}{(\Delta_\text{r}-\omega_\text{b})^2+\kappa_\text{r}^2/4}.
\end{align}
If we ramp the mechanical damping according the expression $\Gamma_\text{r}^\text{opt}(t) = \eta t$, we recover the expression for a time lens \cite{ref:supp_Patera2017TimeLens},
\begin{align}
\langle \hat{b}_\text{s}^\dagger(t)\hat{b}_\text{s}(t) \rangle & = \langle \hat{b_\text{s}}^\dagger(0)\hat{b_\text{s}}(0) \rangle e^{-(\Gamma_\text{b}t+\eta t^2)}.
\end{align}

%

\end{document}